\documentstyle[12pt,preprint,aps,epsf,floats]{revtex}
\def\NPB#1#2#3{Nucl. Phys. {\bf B#1}, #3 (19#2)}

\def\PLB#1#2#3{Phys. Lett. {\bf B#1}, #3 (19#2)}

\def\PRD#1#2#3{Phys. Rev. {\bf D#1}, #3 (19#2)}
\def\PRL#1#2#3{Phys. Rev. Lett. {\bf#1}, #3 (19#2)}
\def\PRep#1#2#3{Phys. Rep. {\bf#1}, #3 (19#2)}

\def\ZPC#1#2#3{Z. Phys. {\bf C#1}, #3 (19#2)}

\newcommand{\postscript}[2]{\setlength{\epsfxsize}{#2\hsize}
   \centerline{\epsfbox{#1}}}

\newcommand{\talpha}{\tilde{\alpha}}

\newcommand{\mheavy}{m_{\rm heavy}}
\newcommand{\mlight}{m_{\rm light}}
\newcommand{\boldm}{\mbox{\boldmath $m$}}
\newcommand{\bolde}{\mbox{\boldmath $e$}}
\newcommand{\boldN}{\mbox{\boldmath $N$}}

\newcommand{\tev}{\text{ TeV}}
\newcommand{\gev}{\text{ GeV}}

\tighten

\begin{document}
\preprint{
\noindent
\begin{minipage}[t]{3in}
\begin{flushleft}
November 1999 \\
\end{flushleft}
\end{minipage}
\hfill
\begin{minipage}[t]{3in}
\begin{flushright}
IASSNS--HEP--99--95\\
MIT--CTP--2907\\
MADPH--99--1139\\
hep-ph/9911255\\
\vspace*{.7in}
\end{flushright}
\end{minipage}
}

\title{
Superheavy Supersymmetry from \\
Scalar Mass--$\boldmath{A}$ Parameter Fixed Points
}

\author{Jonathan A. Bagger,$^a$ Jonathan L. Feng,$^b$ 
Nir Polonsky,$^c$ Ren-Jie Zhang$^d$
\vspace*{.2in}
}
\address{${}^{a}$Department of Physics and Astronomy,
Johns Hopkins University\\ Baltimore, MD 21218  USA}
\address{${}^{b}$School of Natural Sciences,
Institute for Advanced Study\\ Princeton, NJ 08540 USA}
\address{${}^{c}$Center for Theoretical Physics, Massachusetts 
Institute of Technology\\ Cambridge, MA 02139 USA}
\address{${}^{d}$Department of Physics, University of Wisconsin\\
Madison, WI 53706 USA
\vspace*{.2in}
}

\maketitle

\begin{abstract}
In supersymmetric models, the well-known tension between naturalness
and experimental constraints is relieved if the squarks and sleptons
of the first two generations are superheavy, with masses $\mheavy \agt
10 \tev$, and all other superpartners are light, with masses $\mlight
\alt 1 \tev$.  We show that even if all scalar masses and trilinear
$A$ parameters are of order $\mheavy$ at some high scale, a hierarchy
of $\mheavy^2/\mlight^2 \sim 400$ may be generated dynamically through
renormalization group evolution.  The required high energy relations
are consistent with grand unification, or, alternatively, may be
realized in moduli-dominated supersymmetry-breaking scenarios.

\end{abstract}


\section{Introduction}
\label{sec:introduction}

Supersymmetry is a well-motivated framework for extending the standard
model of strong and electroweak interactions~\cite{rep}.  Among its
many virtues, weak-scale supersymmetry provides a natural solution to
the gauge hierarchy problem, realizes gauge unification without the
{\em ad hoc} introduction of additional particles, and elegantly
explains electroweak symmetry breaking in terms of the large top quark
mass.

Given the most general possible set of soft supersymmetry-breaking
terms, however, supersymmetric models violate many well-known
laboratory constraints, particularly those on flavor-changing neutral
currents (FCNC) and CP violation.  Indeed, studies of the
supersymmetric contributions to rare processes place severe
constraints on flavor mixing in the sfermion ($\tilde{f}$) mass
matrices,
\begin{equation}
{m_{f}^2}_{ij}\tilde{f}^*_i \tilde{f}_j\  +\ \text{h.c.}\ ,\qquad
\tilde{f} = Q,U,D,L,E \ ,
\end{equation}
where $i,j$ are generational indices, $Q$ denotes quark SU(2)
doublets, $U$ and $D$ up- and down-type quark singlets, $L$ lepton
doublets, and $E$ lepton singlets.  For example, in the basis in which
the fermion mass matrices are diagonal, the $K_L - K_S$ mass
difference requires
\begin{equation}
\left[{10~{\rm TeV}\over{m}}\right]^2 \left[{{\rm
Re}({m_{Q,D}^2}_{12}/{m^2}) \over 0.1}\right]^2 \alt 1 \ ,
\end{equation}
where $m$ denotes the average squark mass~\cite{constraints}.  If we
assume at most moderate suppressions of the second bracketed term from
squark degeneracy or squark-quark alignment, this bound implies $m
\agt 10 \tev$.  The constraint from the CP-violating parameter
$\epsilon_K$ is even more severe, requiring $m \agt 100 \tev$ for
${\cal{O}}(1)$ CP-violating phases.  Electron and neutron electric
dipole moments provide constraints that are less stringent, but
nevertheless important, because they are flavor-conserving and
therefore unsuppressed by sfermion degeneracy. For ${\cal{O}}(1)$
CP-violating phases, the electric dipole moments require $m \agt 2
\tev$.  Other difficulties, such as too rapid proton decay through
dimension-five operators in grand unified theories~\cite{proton} and
cosmological problems caused by late-decaying moduli~\cite{Polonyi},
are also alleviated if the superpartner and gravitino masses are
heavy.

The above constraints are most easily satisfied if all supersymmetric
scalars are at the 10 TeV scale.  However, such heavy sfermions are in
apparent conflict with naturalness, the requirement that there be no
large cancellations in radiative corrections.  Naturalness suggests
that the superpartner masses should be at most $\sim 1$ TeV.  This
conflict may be avoided, however, by
observing~\cite{drees,Kagan,dgpt,moreminimal} that the most stringent
laboratory constraints apply to quantities associated with the first
two generations, while naturalness primarily restricts the
third-generation sfermions, which couple to the Higgs sector with
large Yukawa couplings.  The laboratory and naturalness constraints
may be simultaneously satisfied if the scalar masses exhibit an
inverted mass hierarchy: third-generation scalars are light with
masses $\mlight\alt 1$ TeV, while the first two generation scalars are
heavy, with masses $\mheavy \agt 10$ TeV.

In this letter we will present a mechanism for generating such an
inverted mass hierarchy dynamically, through renormalization group
evolution.  Our scenario preserves and utilizes the appealing features
mentioned above.  As will become clear, the scenario is compatible
with grand unification, and assumes no additional gauge dynamics or
particle content (aside from a right-handed neutrino).  In addition,
the large top quark Yukawa coupling, which drives electroweak symmetry
breaking, is also used to generate the inverted hierarchy.  In our
scenario, the inverted hierarchy is no accident: both heavy fermions
and light scalars are necessarily associated with large Yukawa
couplings.

This approach was first investigated in Ref.~\cite{fkp}, and then
extended and generalized to the case of unified theories in
Ref.~\cite{BFP}.  Both of these papers assumed the hierarchy $m \gg A,
m_{1/2}$ between the scalar masses, $m$, on the one hand, and the
trilinear couplings $A$ and gaugino masses, $m_{1/2}$, on the other,
as would follow, for example, from an approximate U(1)$_R$ symmetry.
It was found that the third-generation scalar masses can indeed be
exponentially suppressed when renormalization group evolution drives
them to a zero-value infrared fixed point.  In Ref.~\cite{BFP},
hierarchies of $\mheavy^2/\mlight^2 \sim 20$ were realized.  This
solves some of the phenomenological problems discussed previously, but
only alleviates others.  (See Ref.~\cite{BFP} for a detailed
discussion.)

In this study, we relax the restriction on the $A$ parameters and
consider the hierarchy $m, A \gg m_{1/2}$.  Such a hierarchy emerges
naturally, for example, in moduli-dominated supersymmetry-breaking
scenarios~\cite{IL,Brignole,LK}.  The fixed-point structure may be
analyzed as before, but now including both scalar masses and $A$
parameters.  We find that the presence of the $A$ parameters leads to
greatly improved results, with possible hierarchies of $\mheavy^2/
\mlight^2 \sim 400$ and $\mheavy \agt 10 \tev$ naturally achieved.

\section{Naturalness and fixed points}
\label{sec:FP}

The radiative generation of electroweak symmetry breaking is one of
the most important and appealing features of supersymmetry.  At the
weak scale, the necessary condition is conveniently expressed by the
following (tree-level) equation:
\begin{equation}
\frac{1}{2} m_Z^2\ =\ \frac{m_{H_d}^2 - m_{H_u}^2 \tan^2\beta }
{\tan^2\beta -1} - \mu^2,
\label{finetune}
\end{equation}
where $m_{H_d}$ and $m_{H_u}$ are soft Higgs boson parameters, $\mu$
is the supersymmetry-preserving Higgs mass, and $\tan\beta =\langle
H_u^0 \rangle /\langle H_d^0\rangle$ is the usual ratio of Higgs
vacuum expectation values.  A theory is regarded as natural if
Eq.~(\ref{finetune}) is free from large cancellations.

The parameters $m_{H_u}$ and $m_{H_d}$ are determined by
renormalization group evolution from high energies.  Schematically,
the one-loop evolution equations for the soft masses are of the form
\begin{equation}
\dot{m}^2 \sim -Y m^2 + \talpha m_{1/2}^2 - Y A^2 \ ,
\label{scalarRGE}
\end{equation}
where (positive) numerical coefficients are suppressed.  We have used
the following notation: $\dot{(\ )} \equiv d/dt$, where $t \equiv \ln
\left(Q_0^2/Q^2\right)$ and $Q_0$ is the initial renormalization
scale; and $\talpha \equiv g^2/16 \pi^2 = \alpha/4 \pi$ and $Y \equiv
h^2/16 \pi^2$, where $g$ and $h$ are gauge and Yukawa couplings,
respectively.  Our conventions for the $A$ parameters are that the
trilinear soft terms have the form $Ah \phi_i \phi_j \phi_k$. Clearly,
the weak-scale Higgs boson parameters are most strongly affected by
the third-generation sfermions because of their large Yukawa
couplings.  In contrast, the contributions from the first two
generations of sfermions are suppressed by their Yukawa couplings, and
are very small.\footnote{Sfermion masses of the first two generations
may be important if the trace of hypercharge times $m^{2}$ does not
vanish or if their masses are $\agt 20$ TeV. In the latter case, their
two-loop effects are generally not negligible\cite{dgpt,murayama}.
See, however, Ref.~\cite{Nomura} for models in which this difficulty
is avoided.}  From this we conclude that the third generation is
intimately related to the naturalness problem, while the first two
generations effectively decouple.

Let us now assume $m,~A\sim\mheavy$ at some high energy unification
scale, and that the third-generation Yukawa couplings are unified
throughout their renormalization group evolution. (Splittings of the
Yukawa couplings will be taken into account in the numerical results
to follow.)  Let us also assume the gaugino masses are of order
$\mlight$, so they can be neglected in the following analysis. (Note
that gaugino masses cannot be hierarchically suppressed:
$m_{1/2}/\alpha$ is an evolution invariant at one-loop, which, along
with Eq.~(\ref{scalarRGE}), constrains $m_{1/2} \sim \mlight$.)  The
one-loop evolution equations for the third-generation scalar masses
and $A$ parameters may be collectively written as
\begin{equation}
\dot{\boldm}_i^2 = - Y \boldN_{ij} \boldm_j^2 \ ,
\end{equation}
where $\boldm^2$ is a vector containing both $m^2$ and $A^2$
parameters, and $\boldN$ is a matrix of positive constants determined
by color and SU(2) factors.  This set of equations is easily solved by
decomposing arbitrary initial conditions into eigenvectors of
$\boldN$, each of which then evolves independently.  Indeed, if
$\boldN$ has eigenvectors $\bolde_i$ with eigenvalues $\lambda_i$, the
initial condition
\begin{equation}
\boldm^2(t=0) = \sum c_i \bolde_i
\end{equation}
evolves to
\begin{equation}
\boldm^2(t=t_f) = \sum c_i\bolde_i \, {\rm exp}
\left[ - \lambda_i \int_0^{t_f} Y dt\right] \ .
\label{mend}
\end{equation}
The existence of zero-value infrared fixed points follows immediately
from the observation that eigenvectors with large $\lambda_i$ are
asymptotically and exponentially ``crunched'' to zero at the low
scale.  If the initial conditions are dominated by such eigenvectors,
the soft masses are rapidly suppressed relative to their initial
values.  Note that this suppression does not hold in the case of the
first two generations because of the size of their Yukawa
couplings. The dynamically generated hierarchy relates the observed
hierarchy in the fermion sector to the desired (and inverted)
hierarchy in the scalar sector, with large (small) Yukawa couplings
producing heavy (light) fermions and light (heavy) superpartners.  The
exact hierarchy depends on various details such as the initial
conditions for the soft parameters, the evolution interval, and the
Yukawa couplings and their evolution.  Below, we will define a
suppression factor, or crunch factor, $S$, which will serve as a
quantitative measure of the hierarchy achieved for given parameters in
a particular model.

\section{The scalar mass hierarchy}
\label{sec:mssm}

In this section, we consider a model with the particle content of the
minimal supersymmetric standard model with a (heavy) right-handed
neutrino, $N$, as suggested by unified models based on SO(10) or
larger groups.  The analysis is similar, in principle, to that of
Ref.~\cite{BFP}, but is extended to include heavy $A$ parameters; this
is crucial to improving the results.

The superpotential is given by
\begin{equation}
W = -h_t H_u Q U + h_b H_d Q D + h_{\tau} H_d L E - h_n H_u L N \ ,
\end{equation}
where the matter fields $Q$, $U$, $D$, $L$, $E$, and $N$ are those of
the third generation, $H_{u}$ and $H_{d}$ are the two Higgs doublets,
and all other Yukawa couplings may be neglected for this analysis.
The sfermion mass matrices are assumed to be diagonal (see
Sec.~\ref{sec:highenergy}), but not necessarily degenerate at the
boundary, which we take to be the unification scale, $M_G$.  In this
approximation, the evolution equations for the soft parameters are
given by
\begin{eqnarray}
\dot{m}^2_{H_u} &=& - 3 Y_t (m_{H_u}^2 + m^2_U+m^2_Q+A_t^2)
-Y_n (m_{H_u}^2 + m^2_L+m^2_N+A^2_n)\ ,\nonumber\\
\dot{m}^2_{H_d} &=& - 3 Y_b (m_{H_d}^2 + m^2_D+m^2_Q+A_b^2)
-Y_\tau (m_{H_d}^2  + m^2_L+m^2_E+A^2_\tau)\ ,\nonumber\\
\dot{m}^2_{Q} &=& -Y_t (m_{H_u}^2 + m^2_U+m^2_Q+A_t^2)
-Y_b (m^2_{H_d}+m^2_D+m^2_Q+A^2_b)\ ,\nonumber\\
\dot{m}^2_{U} &=& -2Y_t (m_{H_u}^2 + m^2_U+m^2_Q+A_t^2)\ ,\nonumber\\
\dot{m}^2_{D} &=& -2Y_b (m^2_{H_d}+m^2_D+m^2_Q+A^2_b)\ ,\nonumber\\
\dot{m}^2_{L} &=& -Y_n (m_{H_u}^2+m^2_L+m^2_N+A^2_n)
-Y_\tau(m_{H_d}^2+m^2_L+m^2_E+A^2_\tau)\ ,\nonumber\\
\dot{m}^2_{E} &=& -2Y_\tau(m_{H_d}^2+m^2_L+m^2_E+A^2_\tau)\ ,\nonumber\\
\dot{m}^2_{N} &=& -2Y_n (m_{H_u}^2+m^2_L+m^2_N+A^2_n)\ ,\nonumber\\
\dot{A}_t &=& -6 Y_t A_t-Y_b A_b - Y_n A_n\ ,\nonumber\\
\dot{A}_b &=& -Y_t A_t - 6 Y_b A_b-Y_\tau A_\tau \ ,\nonumber\\
\dot{A}_\tau &=& -3 Y_b A_b - 4 Y_\tau A_\tau - Y_n A_n\ ,\nonumber\\
\dot{A}_n &=& -3Y_t A_t - Y_\tau A_\tau - 4 Y_n A_n\ .
\label{rge}
\end{eqnarray}

Let us assume further that the SO(10) unification relations $Y_t = Y_b
= Y_{\tau} = Y_n = Y$ and $A_t = A_b = A_\tau = A_n = A$ are realized
at $M_G$, and neglect for the moment deviations from these relations
from renormalization group evolution. Then
\begin{equation}
\dot{\boldm}^2 = - Y {\boldN} \boldm^2 \ ,
\end{equation}
where ${\boldm^2} = (m_{H_u}^2, m_{U}^2, m_{Q}^2, m_{D}^2, m_{H_d}^2,
m_{L}^2, m_{E}^2, m_{N}^2, A^2)^T$ and
\begin{equation}
{\boldN}= \left[
\begin{array}{ccccccccc}
4 & 3 & 3 & 0 & 0 & 1 & 0 & 1 & 4 \\
2 & 2 & 2 & 0 & 0 & 0 & 0 & 0 & 2 \\
1 & 1 & 2 & 1 & 1 & 0 & 0 & 0 & 2 \\
0 & 0 & 2 & 2 & 2 & 0 & 0 & 0 & 2 \\
0 & 0 & 3 & 3 & 4 & 1 & 1 & 0 & 4 \\
1 & 0 & 0 & 0 & 1 & 2 & 1 & 1 & 2 \\
0 & 0 & 0 & 0 & 2 & 2 & 2 & 0 & 2 \\
2 & 0 & 0 & 0 & 0 & 2 & 0 & 2 & 2 \\
0 & 0 & 0 & 0 & 0 & 0 & 0 & 0 & 16
\end{array} \right] \ .
\end{equation}
The eigenvectors of $\boldN$, and their associated eigenvalues are
\begin{eqnarray}
\bolde_1 : && 16, (2,1,1,1,2,1,1,1,4)\ ,\nonumber\\
\bolde_2 : && 8, ~(2,1,1,1,2,1,1,1,0)\ , \nonumber \\
\bolde_3 : && 6, ~(2,1,0,-1,-2,0,-1,1,0)\ , \nonumber \\
\bolde_4 : && 4, ~(0,1,1,1,0,-3,-3,-3,0)\ , \nonumber \\
\bolde_5 : && 2, ~(0,-1,0,1,0,0,-3,3,0) \ ,\label{evec}
\end{eqnarray}
along with four eigenvectors of zero eigenvalue.  Components of the
initial conditions along $\bolde_1$ and, to a lesser extent,
$\bolde_2$, $\bolde_3$, $\bolde_4$, and $\bolde_5$, are rapidly
suppressed during the renormalization group evolution.  Note that the
eigenvectors $\bolde_1$ and $\bolde_2$ are both consistent with
minimal SO(10) unification, where ${\bf{16}}=\{Q,U,D,L,E,N\}$,
${\bf{10}}\supset\{H_u, H_d\}$, and the SO(10) group is broken in one
step to the standard model group.  The remarkably simple boundary
condition given by $\bolde_1$,
\begin{equation}
4 m_{\bf{16}}^2 = 2 m_{\bf{10}}^2 = A^2 \ ,
\label{mssmbc}
\end{equation}
also has the largest eigenvalue, 16. (For comparison, the largest
eigenvalue in \cite{BFP} was only 8.)  This eigenvector gives rise to
the largest hierarchy.

The above argument neglects many details.  We therefore evaluate the
renormalization group equations numerically at one-loop.  The
independent evolutions of the Yukawa couplings are included, and the
right-handed neutrino is decoupled at its mass, $M_N$.  We do not
include two-loop contributions in our analysis, nor do we include
one-loop terms of order $\mlight$.  Two-loop contributions to the
Yukawa evolution equations were considered in \cite{BFP}, and are
required to establish how large the Yukawa couplings can be before
perturbation theory breaks down. Here we use the results of
\cite{BFP}, and consider only regions of parameter space where
perturbation theory is valid and two-loop effects are small.

We note that two-loop contributions to the scalar mass equations may
be important, at least for the largest hierarchies that we will be
able to achieve ($\mheavy \sim 20 \tev$) \cite{dgpt,murayama}.
However, such effects are not more important than one-loop
${\cal{O}}(\mlight)$ terms that we ignore.  A complete two-loop
analysis requires the specification of all $\mlight$ parameters, and
is highly model-dependent.  Our aim here is only to establish the
possibility of a very large radiative hierarchy, so we take a
model-independent approach. Interesting issues such as radiative
electroweak symmetry breaking, vacuum stability, and the light
spectrum rely on the model-dependent ${\cal{O}}(\mlight)$ terms, and
are outside the scope of this analysis.

\begin{figure}[tb]
\postscript{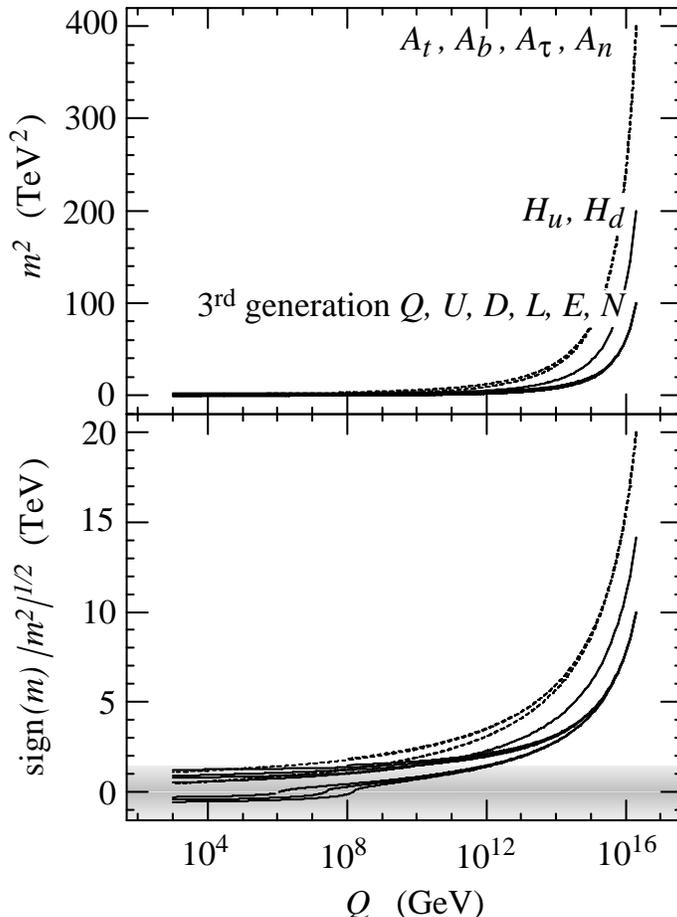}{0.55}
\caption{The renormalization group evolution of the Higgs and
third-generation sfermion masses (solid) and the $A$ term (dotted) in
the supersymmetric standard model with a right-handed neutrino, for
the boundary conditions of Eq.~(\ref{mssmbc}) with $h_G = 2$ and $M_N
= 10^8 \gev$. First- and second-generation scalar masses (not shown)
are approximately evolution invariant.  Model-dependent effects of
order $\mlight^2$ modify solutions in the shaded region, and are not
included.  The suppression factor for this case is $S=330$ (see
text).}
\label{fig:mssmRGE}
\end{figure}

Given our assumptions, the theory is completely specified by the
overall scale, $\mheavy$, and two parameters: $h_G$, the universal
third-generation Yukawa coupling at the unification scale, and $M_N$,
the right-handed neutrino mass.  In Fig.~\ref{fig:mssmRGE}, we show
the evolution of the scalar mass parameters from the unification
scale, $M_G \simeq 2\times 10^{16}$ GeV, to the weak scale, $M_W = 1$
TeV, for the boundary condition of Eq.~(\ref{mssmbc}), with $m_{\bf
16}=10$ TeV, $h_G = 2$ and $M_{N} = 10^{8}\gev$. At the weak scale, an
inverted mass hierarchy is generated, with all third-generation and
Higgs mass parameters $\alt 1 \tev$, and the first two generation
scalar masses (not shown) unsuppressed at $\sim 10\tev$.  The
hierarchy is generated mostly during the first few decades of
evolution, where the Yukawa couplings are large and almost universal,
so the approximations made in the analytical arguments above are
roughly valid.

\begin{figure}[tb]
\postscript{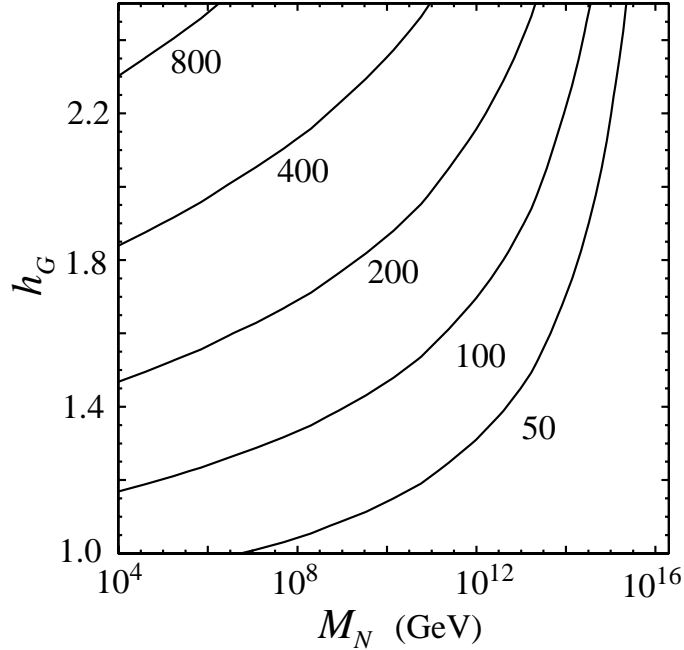}{0.55}
\caption{The suppression factor $S$ for the supersymmetric standard
model with a right-handed neutrino and initial boundary conditions as
given by Eq.~(\ref{mssmbc}).  The parameter $M_N$ is the scale at
which the right-handed neutrino decouples, and $h_G$ is the value of
the universal Yukawa coupling at the unification scale, $M_G \simeq 2
\times 10^{16}\gev$.}
\label{fig:mssmS}
\end{figure}

As in Ref.~\cite{BFP}, we quantify the radiatively generated mass
hierarchy by a suppression, or crunching, factor
\begin{equation}
S \equiv \frac{\bar{m}^2(M_G)}{\bar{m}^2(M_W)} \ ,
\end{equation}
where $\bar{m}^2(Q) \equiv {\rm Av} [| m^2(Q)|]$.  The average is
taken over all scalar degrees of freedom in the theory (but not the
$A$ parameters), properly weighted by color and SU(2) factors.  In
Fig.~\ref{fig:mssmS} we plot $S$ in the $(M_{N}, h_G)$ plane.  We see
that $S$ depends strongly on both of these two parameters. In the
upper left corner, where $h_G\agt 2$ and $M_{N}\alt 10^{8}$ GeV, the
value of $S$ can be as large as 400!  This should be compared with the
maximal crunch factor $S_{\rm max}\sim 20$ in \cite{BFP}.  The large
values of $S$ are a result of the $A$ parameter contributions in the
one-loop evolution equations.  The presence of the $A$ parameters
doubles the maximal eigenvalue, $\lambda_i$, and therefore squares the
maximal $S$.  A crunch factor of $S=400$ corresponds to $\mheavy\sim
20$ TeV.  This can solve all FCNC and CP problems, with the exception
of $\epsilon_K$, which is still beyond bounds if one assumes ${\cal
O}(1)$ phases.

\section{High Energy Scenarios}
\label{sec:highenergy}

The fixed-point mechanism proposed above requires the hierarchy $m, A
\gg m_{1/2}$.  In addition, to achieve the greatest suppression
factors, the scalar mass and $A$ parameters must approximately satisfy
the boundary condition of Eq.~(\ref{mssmbc}). (Small components along
the other eigenvectors are tolerable, especially if their eigenvectors
are also crunched, as are those of Eq.~(\ref{evec}).)  As noted above,
the necessary boundary conditions are consistent with grand
unification. However, they may also find their origin in string
scenarios.  In this section we show that both the required hierarchy
in soft parameters and the specific boundary condition of
Eq.~(\ref{mssmbc}) may be achieved in moduli-dominated
supersymmetry-breaking scenarios with particular assignments of the
modular weights.

The soft supersymmetry-breaking parameters may be viewed as arising
from external spurion fields that parameterize the hidden-sector
supersymmetry breaking.  Generally speaking, soft scalar masses arise
from terms in the K\"ahler potential and therefore depend on $D$-type
supersymmetry breaking.  Gaugino masses and trilinear $A$ parameters
arise from the gauge kinetic function and the superpotential and
depend on $F$-type breaking.  It is certainly possible that different
fields contribute to the different soft terms.  In weakly-coupled
heterotic string theory, for example, the tree-level gauge kinetic
function depends only on the dilaton superfield $S$. The moduli fields
$T$ contribute to the scalar mass terms and the $A$ parameters.  If
supersymmetry breaking is moduli-dominated, with $F_T \gg F_S$, it is
natural to have $m, A \gg m_{1/2}$, as required.\footnote{The $\mu$
and $B\mu$ Higgs mixing parameters may arise from both $F$- and
$D$-type terms, so their origin is more model-dependent.  The
naturalness conditions discussed in Sec.~\ref{sec:FP} require that
they both be ${\cal{O}}(\mlight)$, and we will assume this to be the
case, as a consequence, for example, of an approximate Peccei-Quinn
symmetry.}

\begin{figure}[tb]
\postscript{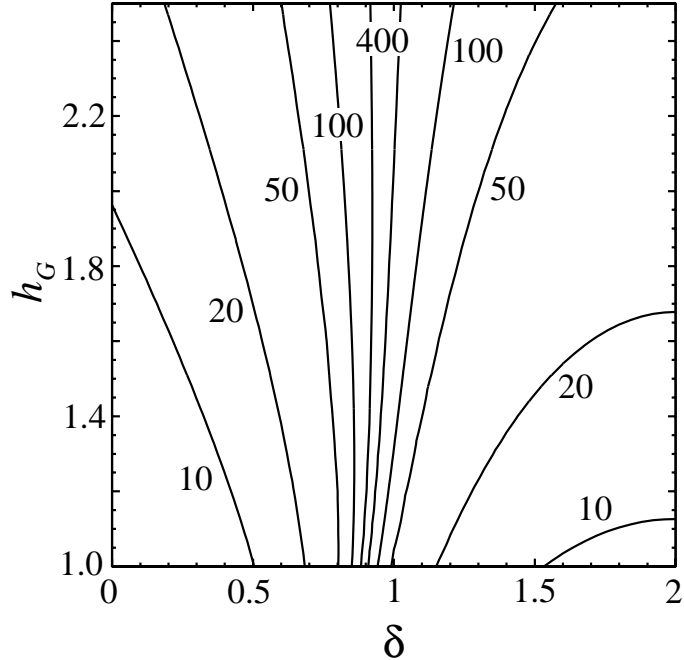}{0.55}
\caption{The suppression factor $S$ for the supersymmetric standard
model with a right-handed neutrino and initial boundary conditions
given by Eq.~(\ref{stringbc}), with $n_{Q, U, D, L, E, N} = 0$, and
$n_{H_u, H_d} = 1$.}
\label{fig:modS}
\end{figure}

In moduli-dominated supersymmetry-breaking scenarios, the scalar
masses are fixed to certain discrete values, determined by
integer-valued modular weights.  It may then be possible not only to
achieve the correct hierarchy, but also the correct ratios of
Eq.~(\ref{mssmbc}).  General discussions and formulae may be found in
Refs.~\cite{IL,Brignole,LK}.  Following Ref.~\cite{Brignole}, for
example, and assuming for simplicity that only one (overall) modulus
$T$ participates in supersymmetry breaking, one finds at tree-level
that
\begin{eqnarray}
m^2_i &=& (1 + n_i) \ m^2_{3/2} \nonumber \\
A_{ijk} &=& [3+ n_i + n_j + n_k - 2\delta(T)] \ m_{3/2} \ ,
\label{stringbc}
\end{eqnarray}
where $n_i$ is the modular weight of field $i$, and $\delta(T) ={\rm
Re}T\ \partial_T \ln h_{ijk}$ is a quantity of order one if the vacuum
expectation value of $T$ and the derivative are of order one in Planck
units. (In this case, the one-loop correction to the gaugino mass can
be at most $\sim\mlight$.)  We find, then, that the correct ratios in
Eq.~(\ref{mssmbc}) are obtained for $n_{Q, U, D, L, E, N} = 0$,
$n_{H_u, H_d} = 1$, and $\delta \sim 1$.  (This choice also implies a
modular weight of $-4$ for the third-generation Yukawa couplings.)
The dependence of $S$ on $h_G$ and $\delta$ is given in
Fig.~\ref{fig:modS}.  While non-negative weights are not
characteristic of the most well-studied examples, such weight
assignments for the standard model fields may be found in Abelian
orbifold models~\cite{IL}.

Two comments are in order before concluding.  Up to this point, we
have neglected the off-diagonal scalar masses.  Given the large
radiative hierarchy, large 1-2 mixings are allowed by all
flavor-changing constraints, although the requirement from
$\epsilon_K$ may impose some restriction if CP-violating phases are
${\cal O}(1)$.  The 1-3 and 2-3 elements are still bounded, however,
by the requirement that the weak-scale theory be tachyon-free.  (See
\cite{BFP} for a more complete discussion of these constraints.)  In
the string context, it may be that these dangerous mixings are
suppressed because different generations have discrete quantum numbers
that forbid off-diagonal terms in the K\"ahler potential~\cite{IL}.
Alternatively, string models often possess additional U(1) groups, and
the quantum numbers of different generations may suppress the scalar
mass mixings~\cite{IL}.  (Such additional groups, however, may make it
more difficult to obtain positive modular weights in the massless
spectrum.)  Note that while such stringy suppressions may already
suppress flavor-violating effects, the superheavy radiative hierarchy
possesses additional virtues, in that the 1-2 mixing need not be
suppressed, and even flavor-conserving difficulties, such as the
electric dipole moments and the Polonyi problem, are solved.

Finally, we note that any tree-level relation such as
Eq.~(\ref{stringbc}) receives quadratically divergent one-loop
corrections from non-renormalizable couplings~\cite{QD}, which
fractionally are of order $\Lambda^2/16 \pi^2$, where $\Lambda$ is the
cut-off in Planck units.  These corrections are not calculable and can
conceivably be large.  Since they are loop-suppressed, we expect them
to be small, of order a few percent, in which case they do not affect
our discussion.

\section{Conclusions}
\label{sec:conclusions}

In this paper we have examined the possibility that all scalar masses
and $A$ parameters are of order some superheavy scale, $\mheavy$, when
they are generated.  The third-generation sfermions are driven to a
scale $\mlight \alt 1$ TeV by their large Yukawa couplings.  For
boundary conditions that we have identified, an attractive zero-value
fixed point allows hierarchies of $\mheavy^2/\mlight^2 \sim 400$.  The
necessary boundary conditions are consistent with SO(10) and similar
grand unified theories.  It is also possible that they have a stringy
origin in terms of modular weights in moduli-dominated
supersymmetry-breaking scenarios.

If we take $\mheavy \sim 20 \tev$, such a scenario leads to a
supersymmetric theory which naturally preserves the gauge hierarchy.
It also solves many difficulties common to supersymmetric models that
follow from flavor- and CP-violating constraints, proton
decay,\footnote{More specifically, proton decay mediated by first- and
second-generation sfermions is highly suppressed; however, recently
analyzed processes involving third-generation scalars may still be
dangerous\cite{proton}.}  and cosmology.  Detailed study of sub-TeV
issues, such as electroweak symmetry breaking and the low energy
spectrum, requires a model-dependent two-loop analysis.

Although we have considered renormalization group evolution below the
unification scale, our observations hold more generally.  For example,
we have verified that the crunch factors found in \cite{BFP}, in the
case of evolution between the Planck and unification scales, are also
enhanced by an order of magnitude once the trilinear $A$ parameters
are included, just as in the case of evolution below the unification
scale discussed here.

Finally, we note that while this scenario suggests that the scalars of
the first and second generation are all well beyond the direct
discovery reaches of planned colliders, these theories are not exempt
from experimental probes.  In particular, this scenario predicts that
all gauginos and third-generation scalars are below $\sim 1$ TeV, so
measurements of the non-decoupling superoblique parameters\cite{so}
should be possible.  The superoblique parameters provide indirect
measurements of $\mheavy$, or the gravitino mass scale, and hence of
the supersymmetry-breaking scale in the hidden sector.  While some
particles may evade direct detection in hierarchical scenarios,
indirect signatures can provide important confirmation and extensive
insights regarding the high energy theory.

\acknowledgements

We thank Kiwoon Choi for discussions concerning quadratically
divergent corrections.  J.A.B. is supported by the U.S.\ National
Science Foundation, grants NSF--PHY--9404057 and NSF--PHY--9970781.
J.L.F. is supported by the Department of Energy (DOE) under contract
No.~DE--FG02--90ER4054442 and through the generosity of Frank and
Peggy Taplin.  N.P. is supported by the DOE under cooperative research
agreement No.~DE--FC02--94ER40818.  R.J.Z. is supported in part by a
DOE grant No.~DE-FG02-95ER40896 and in part by the Wisconsin Alumni
Research Foundation. The authors thank the Aspen Center for Physics,
where this work was initiated, for its hospitality.

\end{document}